\documentclass[conference,a4paper]{IEEEtran}
\usepackage{bm}
\usepackage{algorithm}
\usepackage{algorithmic}
\usepackage{amsfonts}
\usepackage{amssymb}
\usepackage{graphicx}
\usepackage{cite}
\usepackage{subfigure}
\usepackage{epsfig}
\usepackage{url}
\usepackage{bm}
\newtheorem{myth}{Theorem}
\newtheorem{mypro}{Proposition}

\newcommand{\ds}{\displaystyle}

\newcommand{\beqn}{\begin{equation}}
\newcommand{\eeqn}{\end{equation}}
\newcommand{\beqa}{\begin{eqnarray}}
\newcommand{\eeqa}{\end{eqnarray}}
\newcommand{\beqas}{\begin{eqnarray*}}
\newcommand{\eeqas}{\end{eqnarray*}}

\newtheorem{ex}{Example}\newcommand{\Ex}{\begin{ex}\rm}\newcommand{\eex}{\end{ex}}
\newtheorem{re}{Remark}

\newcommand{\Rem}{\begin{re}\rm}
\newcommand{\ere}{\end{re}}

\newcommand{\diag}{{\sf diag}}

\newcommand{\tr}{{\sf Trace}}
\newcommand{\bx}{\bm{x}}
\newcommand{\by}{\bm{y}}

\newcommand{\bX}{\mathbf{X}}
\newcommand{\bY}{\mathbf{Y}}
\newcommand{\bZ}{\mathbf{Z}}
\newcommand{\bz}{\mathbf{z}}
\newcommand{\bC}{\mathbf{C}}
\newcommand{\bH}{\mathbf{H}}
\newcommand{\balpha}{\pmb{\alpha}}
\newcommand{\bbeta}{\pmb{\beta}}

\newcommand{\clN}{{\cal N}}

\newcommand{\bPsi}{\mathbf{\Psi}}
\newcommand{\bPhi}{\mathbf{\Phi}}

\usepackage{amsbsy}
\begin{document}
\title{Two-hop Power-Relaying for Linear Wireless Sensor Networks}

\author{\IEEEauthorblockN{Johann A. Bengua,  Hoang D. Tuan and Ho N. Phien}
\IEEEauthorblockA{Faculty of Engineering and Information Technology\\
University of Technology Sydney\\
Ultimo, Australia\\
Email: johann.a.bengua@student.uts.edu.au\\
tuan.hoang@uts.edu.au\\
ngocphien.ho@uts.edu.au
}
\and
\IEEEauthorblockN{Ha H. Kha}
\IEEEauthorblockA{Faculty of Electrical and Electronics Engineering\\
Ho Chi Minh City University of Technology\\
Ho Chi Minh City, Vietnam\\
Mobile: +84-917330706\\
Email: hhkha@hcmut.edu.vn
}}

\maketitle
\begin{abstract}
This paper presents two-hop relay gain-scheduling control in a Wireless Sensor Network to estimate  a static
target prior characterized by Gaussian  probability distribution. The target is observed by a network of linear
sensors, whose observations are transmitted to a fusion center for carrying out final estimation via a
amplify-and-forward relay node.  We are concerned with the joint transmission power allocation for sensors and relay  to
optimize the  minimum mean square error (MMSE) estimator, which is deployed at the fusion center. Particularly, such highly nonlinear optimization problems are solved by an iterative procedure
of very low computational complexity. Simulations are provided to support the efficiency of our proposed power allocation.
\end{abstract}
\IEEEpeerreviewmaketitle
\section{Introduction}
Wireless Sensor Network (WSN) is an emerging technology that plays a key role
in many applications such as  process monitoring in industrial plants, navigational and guidance systems, radar tracking, sonar ranging, environment monitoring, battlefield surveillance, health care and home automation
\cite{arik10,maj09,nguye8,kar2009,ak02,su08,lian10,ho08}.  Usually the sensors
are geographically distributed to operate in an amplify-and-forward mode \cite{GV03,GV05}. Through wireless communication channels,
the sensors send their own local measurement of a target to a central system, known as the fusion center (FC). The FC filters
these local measurement for a global estimate of the target. The prior knowledge on the target is often assumed to be
Gaussian, in which case the minimum mean square error (MMSE) estimator is defined via the first and second order statistical
moments (the mean vector and covariance matrix) of the jointly Gaussian distributed source and observation \cite[p. 28]{and79}.
As the sensors consume a certain power in transmitting  their observations
to the FC, the sensor power allocation to minimize estimate distortion at the FC has been a subject of
considerable interest \cite{shu07,tha08,dharm8,Jun09,rash12}.  Provided that the target is modelled by a Gaussian random variable, \cite{rash12} shows that the globally optimal distributed Bayes filtering for a linear sensor network (LSN) is computationally tractable by
(convex) semi-definite programming (SDP).  Meanwhile, it is known \cite{R96} that the wireless channels
are Rayleigh fading, suffering the path loss that is proportional to the physical link distance. Therefore the assumption
on the strong wireless channels between the sensors and the FC in all previous works
\cite{shu07,tha08,dharm8,Jun09,rash12} implicitly dictates that the FC must be located near the sensors.
Otherwise the sensors need to consume more transmission power to combat against the path loss of the communication, which is impossible due to either the sensor limited hardware capacity or diminishing battery life. Therefore,
as this paper firstly suggests, it is much more sensible to deploy a relay that is able to amplify and forward the local measurements
of the sensors to the FC. Accordingly, the interested problem is to jointly allocate the relay powers and sensor powers to optimize
the MMSE estimator at the FC. Unlike the separated sensor power problem which is convex and solved by SDP \cite{rash12}, this new
joint power control is no longer convex. Nevertheless, we will show in this paper that it can be addressed by successive convex
programs, each of which admits a closed-form solution.\\
The paper is structured as follows. After the Introduction, Section II introduces
the two-hop relayed wireless sensor network and gives the power optimization formulation.
Section III is devoted to its solution by successive convex programming. Section IV
provides a preliminary simulation to support the result of Section V. Section V concludes
the paper. Due to the space limitation, all proofs are ommited.\\
Most of the notations used in the paper are described here. Bold
lower-case and upper-case symbols are used to represent vectors and
matrices respectively. By $\mathbf{A}\succeq \mathbf{B}$ it means
$\mathbf{A}-\mathbf{B}\succeq 0$, i.e. $\mathbf{A}-\mathbf{B}$ is a
positive definite matrix. $\diag[a_i]_1^N$ is a diagonal matrix with ordered diagonal entries
$a_1,a_2,\ldots ,a_N$.
$\sqrt{\mathbf{q}}$ for a vector $\mathbf{q}$ with nonnegative
components is component-wise understood. Trace of a square matrix
$\mathbf{A}$ is expressed by $\tr(\mathbf{A})$. $\mathbb{E}[.]$ is the expectation operator. $\bX\sim
p_\bX(\bx)$ is referred to a random variable (RV) $\bX$ with
probability density function (PDF)  $p_\bX(\bx)$.  $\bm{m}_{\bX}$ is
its expectation $\mathbb{E}[\bX]$, while $\mathbf{C}_{\bm{X}}$ is
its auto-covariance matrix
$\mathbb{E}[(\bm{X}-\bm{m}_{\bm{X}})(\bm{X}-\bm{m}_{\bm{X}})^T]$ and
$\mathbf{C}_{\bm{X}\mathbf{Y}}$ is its cross-covariance matrix
$\mathbb{E}[(\bm{X}-\bm{m}_{\bm{X}})(\mathbf{Y}-\bm{m}_{\mathbf{Y}})^T]$
with another RV $\mathbf{Y}$. Similarly $\mathbf{R}_{\bm{X}}$ is its
auto-correlation matrix
$\mathbb{E}[\bm{X}\bm{X}^T]=\bC_{\bX}+\bm{m}_{\bX}(\bm{m}_{\bX})^T$
and $\mathbf{R}_{\bm{X}\mathbf{Y}}$ is its cross-correlation matrix
$\mathbb{E}[\bm{X}\mathbf{Y}^T]=\bC_{\bX\bY}+\bm{m}_{\bX}(\bm{m}_{\bY})^T$
with another RV $\mathbf{Y}$. $\bX|\bY$ is a random variable $\bX$
restricted by  a realization of the conditioning random variable
$\bY$ and accordingly $\bX|\bY=\by$ is a random variable restricted
by  the value $\bY=\by$ of $\bY$.
$\clN(\bx;\bm{m}_{\bX},\bC_{\bX}):=\frac{1}{\sqrt{2\pi\det(\mathbf{C}_{\bm{X}})}}\exp\left(
-\frac{1}{2}(\bm{x}-\bm{m}_{\bm{X}})^T\mathbf{C}_{\bm{X}}^{-1}(\bm{x}-\bm{m}_{\bm{X}})\right)$
is a Gaussian distribution so $\bX\sim
\clN(\bx;\bm{m}_{\bX},\bC_{\bX})$  means that $\bm{X}$ is Gaussian
RV with expectation $\bm{m}_{\bm{X}}$  and covariance
$\mathbf{C}_{\bm{X}}$.

\section{Behavioral framework based relayed optimization}
Suppose $(\bm{X},\mathbf{Y})$ is a jointly Gaussian  RV characterized by
\begin{equation}\label{1mod}
(\bm{X},\mathbf{Y}) \sim f_{\bX,\bY}(\bx,\by):=\mathcal{N}\left( (\bm{x},\mathbf{y});\bm{m}_{\bX,\bY},\mathbf{C} \right)
\end{equation}
with  $\bm{m}_{\bX,\bY}=\left(\begin{array}{c}\bm{m_X}\cr
\bm{m}_\mathbf{Y}\end{array}\right)$, $\mathbf{C} =
\left(\begin{array}{cc}\mathbf{C}_{\bm{X}} &
\mathbf{C}_{\bm{XY}}\\ \mathbf{C}_{\bm{YX}} &
\mathbf{C}_{\bm{Y}}
\end{array}\right)$.
The MMSE estimate for $\bX$ based on the measurement $\bY=\by$ is
\begin{eqnarray}
\hat{\bm{x}} =\bm{m_X}+\mathbf{C}_{\mathbf{Y}\bm{X}}^T\mathbf{C}_{\bY}^{-1}(\mathbf{y}-\bm{m}_\mathbf{Y})\label{zes4}
\end{eqnarray}
with MSE covariance
\begin{eqnarray}
\bC &=&\ds\int (\bx-\hat{\bm{x}} )(\bx-\hat{\bm{x}} )^Tf_{\bX|\bY=\by}(\bx)d\bx\label{lcov1}\\
&=&\bC_{\bX}-\bC_{\bX\bY}\bC^{-1}_{\bY}\bC_{\bY\bX},\label{lcov2}
\end{eqnarray}
MMSE of Bayesian estimate $\mathbb{E}[||\bX-\hat{\bX}|\bY=y||^2]$ for $\bX$ based  on observation $\bY=\by$ is thus
\begin{equation}\label{ad2}
\epsilon^2  = \tr(\mathbf{C} )
\end{equation}
Now, consider a Gaussian target $\clN(\bm{m}_{\bm{X}},\bm{C}_{\bm{X}})$ in $N$-dimensional space, which is
observed by $M$ spatially distributed linear sensors $\mathbf{Y}$ as
\begin{equation}\label{lsn1}
\bm{y}=\bm{G}\bm{x}+\bm{n},
\end{equation}
where $\bm{n}$ is white noise $\clN(0,\bm{R}_{\bm{n}})$, which
independent from $\bm{x}$. It is easy to see that $(\bm{X},\mathbf{Y})$
constitutes the behavioral equation (\ref{1mod}) with
\begin{equation}\label{lsn2}
\begin{array}{c}
\bm{m}_{\bm{Y}}=\bm{G}\bm{m}_{\bm{X}}, \bm{C}_{\bm{Y}}=\bm{G}\bm{C}_{\bm{X}}\bm{G}^T+\bm{R}_{\bm{n}},\\
\bm{C}_{\bm{XY}}=\bm{C}_{\bm{YX}}^T=\bm{C}_{\bm{X}}\bm{G}^T.
\end{array}
\end{equation}
Accordingly $\mathbf{y}=(y_1, y_2,...,y_M)^T$ is the sensor observations with
\begin{equation}\label{powery}
||y_j||^2=\bC_{\bY}(j,j)+\bm{m}_{\bY}^2(j).
\end{equation}
The sensors send these noise corrupted observations $y_j$ to the relay over wireless flat-fading
time-orthogonal communication channels \cite{shu07}. The signals received at the relay
can thus be written as
\begin{equation}\label{relay1}
z_{jR}=\sqrt{h_{jR}\alpha_j}y_j+w_{jR}, j=1, 2,...,M,
\end{equation}
where $\sqrt{h_{jR}}$ is the channel gain between sensor $j$ and the relay, $w_{jR}$ is a corrupt
noise, which can be assumed white with power $\sigma_{jR}$ and independent with the $y_j$ and
$\sqrt{\alpha_j}$ is to control the transmit power $P_j$ of sensor $j$
\[
P_j=\alpha_j||y_j||^2=(\bC_{\bY}(j,j)+\bm{m}_{\bY}^2(j))\alpha_j,
\]
which is subject to a fixed sum power budget $P_T>0$
\begin{equation}\label{budget}
\ds\sum_{j=1}^MP_j=\ds\sum_{j=1}^M||y_j||^2\alpha_j\leq P_T.
\end{equation}
The relay will then amplify these
received signals $z_{jR}$ to power level $\beta_j$ before forward them to the FC, so the
received signals at the FC are
\begin{eqnarray}
z_j=&\sqrt{h_{jD}}\sqrt{\beta_j/||z_{jR}||^2}z_{jR}+w_{jD}&\nonumber \\
=&\sqrt{h_{jD}h_{jR}\beta_j\alpha_j/(h_{jR}||y_j||^2\alpha_j+\sigma_{jR})}y_j+w_j,&\label{relay2}
\end{eqnarray}
where $\sqrt{h_{jD}}$ is the channel gain between the relay and the FC in the carrier $j$,
$w_{jD}$ is the background noise at the FC, which can be assumed noised with power $\sigma_{jD}$
and independent with $z_{jR}$. Accordingly,
\[
w_j=\sqrt{h_{jD}\beta_j/(h_{jR}||y_j||^2\alpha_j+\sigma_{jR})}w_{jR}+w_{jD}
\]
is white noise with power $h_{jD}\beta_j/(h_{jR}||y_j||^2\alpha_j+\sigma_{jR})+\sigma_{jD}$. The
power levels $\beta_j$ are constrained by the relay power budget $P_R$ as
\begin{equation}\label{rbudget}
\sum_{i=1}^M\beta_j\leq P_R.
\end{equation}
Thus, the signals received at the FC can be written in a vector form by
\begin{equation}\label{2mod}
\mathbf{Z}_{\bm{\alpha},\bm{\beta}}=\mathbf{H}_{\bm{\alpha},\bm{\beta}}\mathbf{Y}+\mathbf{W}_{\bm{\alpha}, \bm{\beta}},\end{equation}
where  $\mathbf{H}_{\bm{\alpha}, \bm{\beta}}\in\mathbb{R}^{M\times M}$ is defined by
 \begin{eqnarray*}\label{rmat}
\mathbf{H}_{\bm{\alpha},\bbeta}=\diag\left[\sqrt{h_{jD}h_{jR}\beta_j\alpha_j/(h_{jR}||y_j||^2\alpha_j+\sigma_{jR})}\right]_1^M,
\end{eqnarray*}
and $\mathbf{W}_{\bm{\alpha},\bm{\beta}}\sim\mathcal{N}(0,\mathbf{C}_{\bm{\alpha},\bm{\beta}})$ with diagonal matrix
\begin{eqnarray*}
\mathbf{C}_{\bm{\alpha},\bm{\beta}}=\diag\left[h_{jD}\beta_j/(h_{jR}||y_j||^2\alpha_j+\sigma_{jR})+\sigma_{jD}\right]_{1}^M
\end{eqnarray*} is the total  noise.  \\
Based on (\ref{1mod}) and (\ref{2mod}), one can write the behavioral equation
\begin{eqnarray}
(\bm{X},\mathbf{Z}_{\balpha,\bbeta}) \sim & f_{\bX,\bZ_{\balpha,\bbeta}}(\bx,\bz)&\nonumber\\
=&\ds\mathcal{N}\left( (\bm{x},\mathbf{z});\left(\begin{array}{c}\bm{m_X}\cr
\bH_{\balpha,\bbeta}\bm{m}_\mathbf{Y})\end{array}\right), \right.&\nonumber\\
&\left.\left(\begin{array}{cc} \mathbf{C}_{\bX}&\bC_{\bX\bY}\bH_{\balpha,\bbeta}\cr
\bH_{\balpha,\bbeta}\bC_{\bY\bX}&\bH_{\balpha,\bbeta}\bC_{\bY}\bH_{\balpha,\bbeta}+\bC_{\balpha,\bbeta}\end{array}\right) \right).&\label{ad3}
\end{eqnarray}
From (\ref{lcov2}), the Bayesian optimal MMSE estimate based on FC output $\mathbf{Z}_{\balpha,\bbeta}=\bz$ is
\begin{equation}\label{Bayesian}
\bm{\hat{x} }\triangleq  \mathbb{E}[\bm{X}|\mathbf{Z}_{\balpha,\bbeta}=\bz]=
  \bm{m}_{\bX|\bZ_{\balpha,\bbeta}}
\end{equation}
where
\begin{equation}\label{ad4}
\begin{array}{lll}
\bm{m}_{\bX|\bZ_{\balpha,\bbeta}}&=&\bm{m}_{\bm{X}}+ \mathbf{C}_{\mathbf{Y}\bm{X}}^{T}\mathbf{H}_{\balpha,\bbeta}(\mathbf{H}_{\balpha,\bbeta}\mathbf{C}_{\bY}\mathbf{H}_{\balpha,\bbeta}
+\mathbf{C}_{\balpha,\bbeta})^{-1}\\
&&\times (\mathbf{z}-\mathbf{H}_{\balpha,\bbeta}\bm{m}_{\bY}).
\end{array}
\end{equation}
Accordingly,
\begin{equation}\label{ad5b}
\{\bX|\bZ_{\balpha,\bbeta}=\bz\} \sim p_{\bX|\bZ_{\balpha,\bbeta}=\bz}=
\clN(\bx,\bm{m}_{\bX|\bZ_{\balpha,\bbeta}},
\bC_{\bX|\bZ_{\balpha,\bbeta}}),
\end{equation}
where
\begin{eqnarray}
\bC_{\bX|\bZ_{\balpha,\bbeta}}&=&\bC_{\bX}-\bC^{T}_{\bY\bX}\bH_{\balpha,\bbeta}(\bH_{\balpha,\bbeta}\bC_{\bY}\bH_{\balpha,\bbeta}
+\bC_{\balpha,\bbeta})^{-1}\nonumber\\
&&\times \bH_{\balpha,\bbeta}\bC_{\bY\bX}\label{ad5a}\\
&=&(\bC_{\bX}-\bC^{T}_{\bY\bX}(\bC_{\bY})^{-1}\bC_{\bY\bX})+\bC^{T}_{\bY\bX}
(\bC_{\bY})^{-1}\nonumber\\
&&\times ((\bC_{\bY})^{-1}+\diag[\varphi_j(\alpha_j,\beta_j)]_1^M)^{-1}\nonumber\\
&&\times (\bC_{\bY})^{-1}
\bC_{\bY\bX} \label{ad5ab}
\end{eqnarray}
with
\begin{equation}\label{fj}
\begin{array}{c}
\varphi_j(\alpha_j,\beta_j)=p_j\ds\frac{\alpha_j\beta_j}{q_j\alpha_j
+r_j\beta_j+\sigma_j},\\
p_j=h_{jR}h_{jD}, q_j=h_{jR}\sigma_{jD}||y_j||^2, \\
r_j=h_{jD}, \sigma_j=\sigma_{jD}\sigma_{jR}.
\end{array}
\end{equation}
\section{Tractable successive convex optimization}
Consider
\begin{equation}\label{lop1}
\min_{\balpha_j\geq 0, \beta_j\geq 0, j=1, 2,...,M}\ \tr(\bC (\balpha,\bbeta))\ :\ (\ref{budget}), (\ref{rbudget}),
\end{equation}
which is equivalent to the following  program
\begin{equation}\label{po}
\begin{array}{c}
\ds\min_{\balpha,\bbeta} \varphi(\balpha,\bbeta):=\tr(\bPsi^H(\bPhi+\diag[\varphi_j(\alpha_j,\beta_j)]_1^M)^{-1}\bPsi)\\
\quad \mbox{subject to}\quad (\ref{budget}), (\ref{rbudget}),\end{array}
\end{equation}
where
\begin{equation}\label{convex1}
\bPsi=\bC^{-1}_{\bY}\bC_{\bY\bX}, \bPhi=\bC^{-1}_{\bY}.
\end{equation}
It can be seen from (\ref{fj}) and (\ref{po}) that (\ref{po}) is a highly nonconvex optimization
in $(\balpha,\bbeta)$. Nevertheless, in what follows we develop a successive procedure, which yields
an optimal (possibly local) solution of (\ref{po}).\\
Given $(\alpha^{(\kappa)},\beta^{(\kappa)})$ we now process the following successive approximations.
Define
\begin{equation}\label{suc1}
\begin{array}{lll}
\varphi_j^{(\kappa)}&=&\varphi_j(\alpha_j^{(\kappa)}, \beta_j^{(\kappa)}),\\
\pmb{\Theta}^{(\kappa)}&=&\diag[\varphi_j^{(\kappa)}]_{1}^M(\bPhi+\diag[\varphi_j^{(\kappa)}]_1^M)^{-1}\bPsi
\bPsi^H(\bPhi\\
&&+\diag[\varphi_j^{(\kappa)}]_1^M)^{-1}\diag[\varphi_j^{(\kappa)}]_{1}^M\\
&\succ& 0,\\
\rho_j^{(\kappa)}&=&\pmb{\Theta}^{(\kappa)}(j,j)>0,
\end{array}
\end{equation}
where $\pmb{\Theta}^{(\kappa)}(j,j)$ is the $j$-th diagonal entry of
$\Theta^{(\kappa)}$.
\begin{myth}\label{succth} The following inequalities hold true for all $\balpha>0$ and $\bbeta>0$,
\begin{eqnarray}
\varphi(\balpha,\bbeta)&\leq&\varphi^{(\kappa)}(\balpha,\bbeta)\label{suc3}
\end{eqnarray}
where
\begin{eqnarray}
\varphi(\balpha,\bbeta)&:=&\varphi(\alpha^{(\kappa)}, \beta^{(\kappa)})+\ds\sum_{j=1}^M\rho_j^{(\kappa)}[\frac{r_j}{p_j\alpha_j}+\frac{q_j}{p_j\beta_j}\nonumber\\
&&+\ds\frac{\sigma_j}{2p_j}(\frac{\alpha_j^{(\kappa)}}{\beta_j^{(\kappa)}\alpha_j^2}
+\frac{\beta_j^{(\kappa)}}{\alpha_j^{(\kappa)}\beta_j^2})-\frac{1}{\varphi_j^{(\kappa)}}].\label{suc3a}
\end{eqnarray}
\end{myth}
Function $\varphi^{(\kappa)}$ is convex majorant of the highly nonconvex function $\varphi$.
According we consider the following majorant minimization
\begin{eqnarray}
\ds\min_{\balpha,\bbeta} \varphi^{(\kappa)}(\balpha,\bbeta)\quad \mbox{subject
to}\quad (\ref{budget}), (\ref{rbudget}).\label{suc5}
\end{eqnarray}
\begin{mypro}\label{succpro}
Whenever $(\alpha^{(\kappa)}, \beta^{(\kappa)})$ is feasible to (\ref{budget}), (\ref{rbudget}), the optimal solution
$(\alpha^{(\kappa+1)}, \beta^{(\kappa+1)})$ of convex program  (\ref{suc5}) is a feasible solution
of nonconvex program (\ref{po}), which is better than $(\alpha^{(\kappa)}, \beta^{(\kappa)})$, i.e.
\begin{equation}\label{suc6}
\varphi(\alpha^{(\kappa+1)}, \beta^{(\kappa+1)})<\varphi(\alpha^{(\kappa)}, \beta^{(\kappa)})
\end{equation}
as far as $(\alpha^{(\kappa+1)}, \beta^{(\kappa+1)})\neq (\alpha^{(\kappa)}, \beta^{(\kappa)})$.
\end{mypro}
We now show that the convex program (\ref{suc5}) admits the optimal solution in closed-form. Indeed, (\ref{suc5}) boils down to
\begin{equation}\label{suc7}
\ds\min_{\balpha,\bbeta}\  \ds\sum_{j=1}^M(\frac{a_j^{(\kappa)}}{\alpha_j}+\frac{b_j^{(\kappa)}}{\beta_j}+
\frac{c_j^{(\kappa)}}{2\alpha_j^2}+
\frac{d_j^{(\kappa)}}{2\beta_j^2})\quad \mbox{subject
to}\quad (\ref{budget}), (\ref{rbudget})
\end{equation}
with
\begin{equation}\label{suc8}
\begin{array}{c}
a_j^{(\kappa)}=\rho_j^{(\kappa)}r_j/p_j, b_j^{(\kappa)}=\rho_j^{(\kappa)}q_j/p_j,\\
 c_j^{(\kappa)}=\rho_j^{(\kappa)}\sigma_j\alpha_j^{(\kappa)}/(p_j\beta_j^{(\kappa)}),\\
 d_j^{(\kappa)}=\rho_j^{(\kappa)}\sigma_j\beta_j^{(\kappa)}/(p_j\alpha_j^{(\kappa)})
 \end{array}
\end{equation}
By using the Lagrangian multiplier method, it can be shown that the
optimal $\alpha_j$ and $\beta_j$  are the unique positive roots of the
following compressed cubic equations
\begin{eqnarray}
a_j^{(\kappa)}\alpha_j+c_j^{(\kappa)}=\lambda_T||y_j||^2\alpha_j^3, j=1, 2,...,M,\label{suc9}\\
b_j^{(\kappa)}\beta_j+d_j^{(\kappa)}=\lambda_R\beta_j^3, j=1,2, ...,M,\label{suc10}
\end{eqnarray}
where $\lambda_T>0$ and $\lambda_R>0$ such that $\alpha_j$ and $\beta_j$ satisfy the power constraints
(\ref{budget}) and (\ref{rbudget}) at equality sign. Accordingly,\footnote{the unique positive root of cubic equation
$ax^3-cx-d=0$ with $a>0$, $c>0$, $d>0$ is $[(d/2a)+\sqrt{(d/2a)^2+(c/3a)^2}]^{1/3}+[(d/2a)-\sqrt{(d/2a)^2+(c/3a)^2}]^{1/3}$}
\begin{equation}\label{suc11}
\begin{array}{c}
\alpha_j^{(\kappa+1)}=\nonumber\\
\ds\left\{\frac{c_j^{(\kappa)}}{2\lambda_T||y_j||^2}+\left[(\frac{c_j^{(\kappa)}}{2\lambda_T||y_j||^2})^2
+(\frac{a_j^{(\kappa)}}{3\lambda_T||y_j||^2})^2 \right]^{1/2}\right\}^{1/3}\\
+\left\{\frac{c_j^{(\kappa)}}{2\lambda_T||y_j||^2}-\left[(\frac{c_j^{(\kappa)}}{2\lambda_T||y_j||^2})^2+
(\frac{a_j^{(\kappa)}}{3\lambda_T||y_j||^2})^2 \right]^{1/2}\right\}^{1/3}
\end{array}
\end{equation}
\begin{eqnarray}
\beta_j^{(\kappa+1)}=&&\ds\left\{\frac{d_j^{(\kappa)}}{2\lambda_R}+\left[(\frac{d_j^{(\kappa)}}{2\lambda_R})^2
+(\frac{b_j^{(\kappa)}}{3\lambda_R})^2 \right]^{1/2}\right\}^{1/3}\nonumber\\
&&+\left\{\frac{d_j^{(\kappa)}}{2\lambda_R}-\left[(\frac{d_j^{(\kappa)}}{2\lambda_R})^2+
(\frac{b_j^{(\kappa)}}{3\lambda_R})^2
\right]^{1/2}\right\}^{1/3},\label{suc12}
\end{eqnarray}
where $\lambda_T>0$ and $\lambda_R$ are chosen so that such
$\alpha_j$ and $\beta_j$ satisfy the power constraints
(\ref{budget}) and (\ref{rbudget}) at equality sign, which can be
located by the following golden search.\\
{\bf Golden search.} {\it Set
$\lambda_{T\min}=\ds\max_{j=1,...,M}[(a_j^{(\kappa)}/P_T^2+c_j^{(\kappa)}/P_T^3)/||y_j||^2]$
 and define $\alpha_j$ by (\ref{suc11}) for
$\lambda_T=2\lambda_{T\min}$. If $\sum_{j=1}^M||y_j||^2\alpha_j>P_T$
set $\lambda_{T\min}=\lambda_T$  and repeat. Otherwise set
$\lambda_{T\max}=\lambda_T$. Restart from
$\lambda_T=(\lambda_{T\min}+\lambda_{T\max})/2$ and define
$\alpha_j$ by (\ref{suc11}). If $\sum_{j=1}^M||y_j||^2\alpha_j>P_T$
reset $\lambda_{T\min}=\lambda_T$. Otherwise reset
$\lambda_{T\max}=\lambda_T$. Process till
$\sum_{j=1}^M||y_j||^2\alpha_j=P_T$. \\
Set
$\lambda_{R\min}=\ds\max_{j=1,...,M}(b_j^{(\kappa)}/P_R^2+d_j^{(\kappa)}/P_R^3)$,
 and define $\beta_j$ by (\ref{suc12}) for
$\lambda_R=2\lambda_{R\min}$. If $\sum_{j=1}^M\beta_j>P_R$ set
$\lambda_{R\min}=\lambda_R$ and repeat. Otherwise set
$\lambda_{R\max}=\lambda_R$. Restart from
$\lambda_R=(\lambda_{R\min}+\lambda_{R\max})/2$ and define $\beta_j$
by (\ref{suc12}). If $\sum_{j=1}^M\beta_j>P_T$ reset
$\lambda_{R\min}=\lambda_R$. Otherwise reset
$\lambda_{R\max}=\lambda_R$. Process till $\sum_{j=1}^M\beta_j=P_T$.
}\\
{\bf ALGORITHM 1.} {\it Initialized from  $(\alpha^{(0)},
\beta^{(0)})$ feasible to (\ref{budget}) and (\ref{rbudget}), for $\kappa=0, 1, ...$
generate a feasible solution $(\alpha^{(\kappa+1)}, \beta^{(\kappa+1)})$ for $\kappa=0, 1, ...,$
according to formula (\ref{suc11}) and  (\ref{suc12}) until
\begin{equation}\label{stop}
\frac{\varphi(\alpha^{(\kappa)}, \beta^{(\kappa)})-\varphi(\alpha^{(\kappa+1)}, \beta^{(\kappa+1)})}
{\varphi(\alpha^{(\kappa)}, \beta^{(\kappa)})}\leq \epsilon
\end{equation}
for a given tolerance $\epsilon$.
}

It follows from Proposition \ref{succpro} that.
\begin{mypro}\label{sucpro1} Algorithm 1 generates a sequence
$\{(\alpha^{(\kappa)}, \beta^{(\kappa)})\}$ of improved solutions,
which converges to an optimal solution of the nonconvex problem
(\ref{po}).
\end{mypro}

\section{Simulations}
The proposed algorithm is validated via two LSN experiments: random scalar targets and random vector targets. In both cases, 10000 Monte Carlo channel realizations are generated and targets are static. The background noise power for all parties (sensors, relay and FC) are assumed to be $\bm{R_n}=\mbox{diag}[\sigma_{jR}]_1^M=\mbox{diag}[\sigma_{jD}]_1^M=\bm{I}$, where $\bm{I}$ is the identity matrix. The channel gains $h_{jR}$ and $h_{jD}$ are determined according to $h=\mbox{SNR}(\lambda/4\pi d)^2$, with the distance
 between two ends $d$, signal wavelength $\lambda$ and signal-to-noise ratio $\mbox{SNR}$. The transmit power budgets $P_T=[0.1, 0.2, \ldots, 1.0]$ and the relay power budget is fixed at $P_R=5$ for both random scalar and vector experiments. Random permutations of sensor placements surrounding the mean of the targets $\bm{m_X}$ are generated for each channel realization.

\begin{figure}[!htbp]
\centering
\includegraphics[width=3.2in]{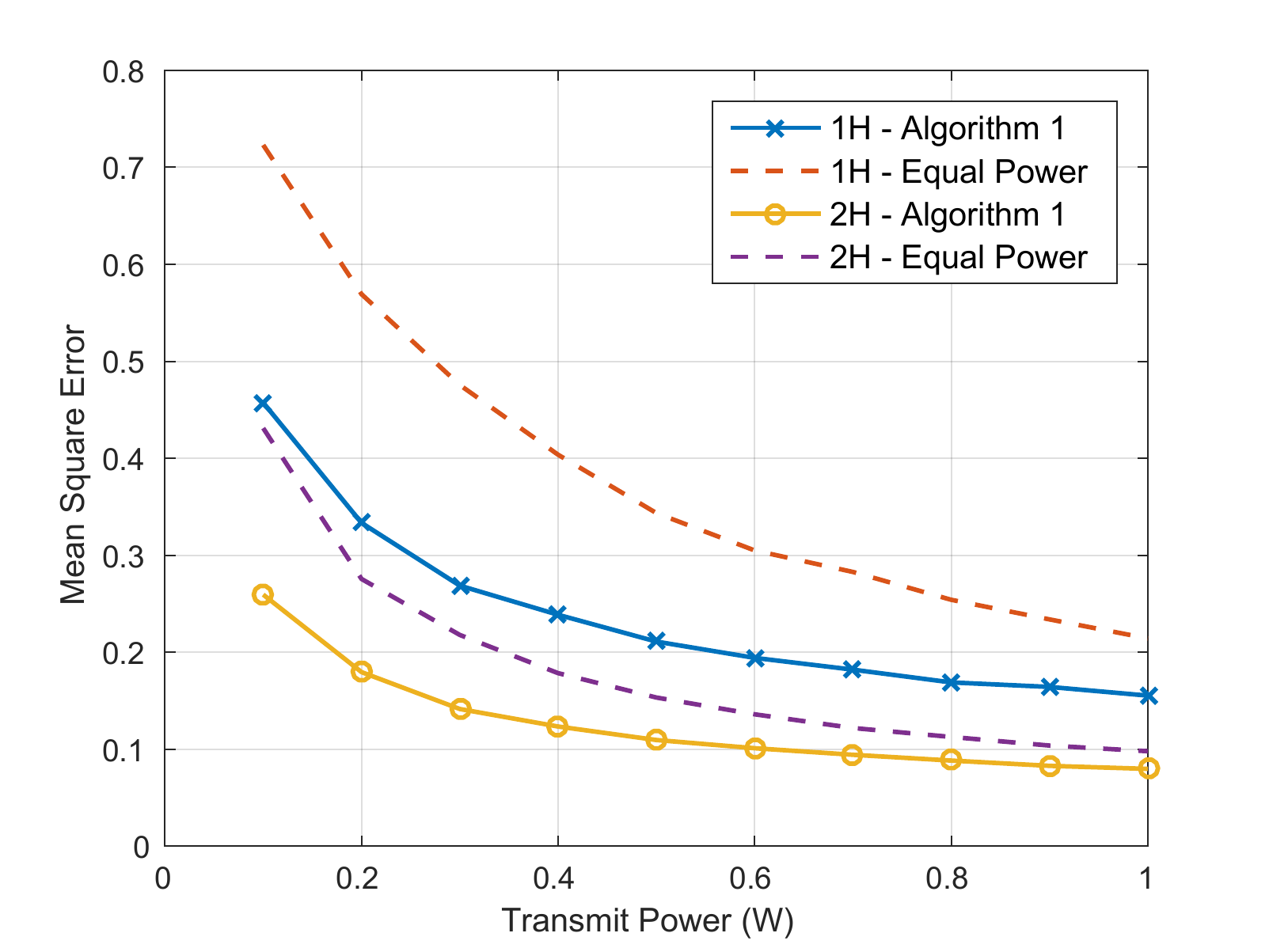}
\caption{MSE versus power budget for random scalar targets. 1H and 2H refers to one-hop and two-hop experimental conditions.}\label{msescale}
\end{figure}

\begin{figure}[!htbp]
\centering
\includegraphics[width=3.2in]{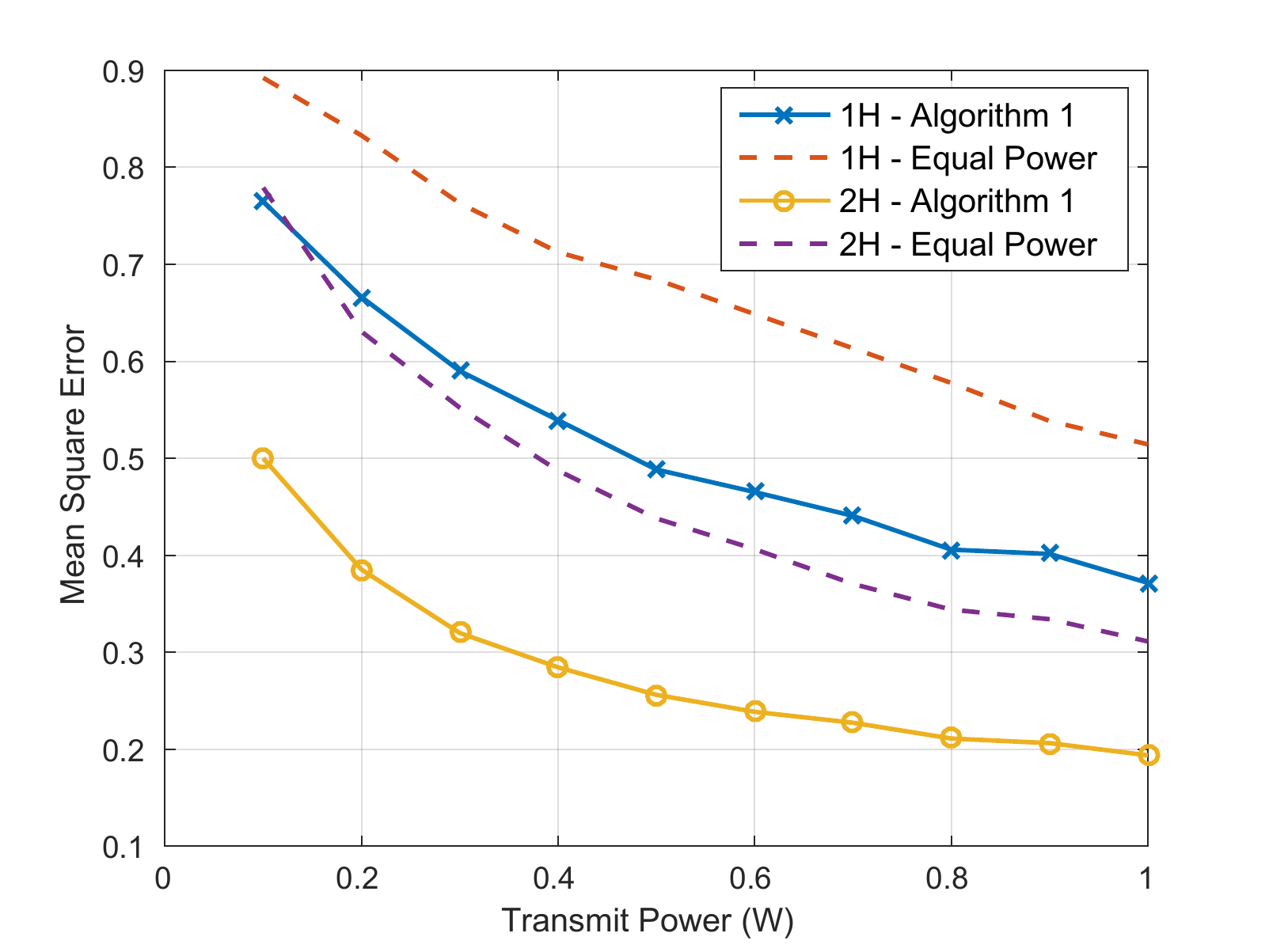}
\caption{MSE versus power budget for random vector targets.}\label{msevect}
\end{figure}

For random scalars it is assumed that ten ($M = 10$) sensors are in different channel conditions
\[
\bm{G}=[1.00,1.11,1.22,1.33,1.44,1.55,1.66,1.77,1.88,2.0]^T.
\]
The mean square error (MSE) results for random scalars are shown in Fig. \ref{msescale}.
In this figure we compare one-hop (sensors communicate directly to the FC, $d = 400m$)\footnote{The one-hop case corresponds to the case
when the relay and FC are the same. The power budget is still kept at $P_T$ to reflect that it cannot be increased
either due to the sensor limited hardware capacity or to save the sensor battery life.}
 and two-hop conditions ($d=200m$ for sensor to relay and relay to FC) as well as their respective uniform power distributions. It can be seen that two-hop is optimal in the majority of power budgets. For random vectors ($N=3$), each sensor node performs range, elevation angle and azimuth measurements
\beqa
\bm{g_j}(\bm{x}) = \nonumber\\
\left(
\begin{array}{c}
 \sqrt{(x(1)-s_{j,x})^2+(x(2)-s_{j,y})^2+(x(3)-s_{j,z})^2}      \\
 (x(2)-s_{j,y})/(x(1)-s_{j,y})      \\
 (x(3)-s_{j,z})/\sqrt{(x(1)-s_{j,x})^2+(x(2)-s_{j,y})^2}   \label{gnlvect}
\end{array}
\right),
\eeqa
with $s_{j\bm{\hat{p}}}$ ($\bm{\hat{p}}=\{x,y,z\}$) being the Cartesian coordinates of a sensor $j$. The power allocation is distributed to all $M = 10$ sensors and the three measuring components. Subsequently, nonlinear maps $\bm{g_j}(\bm{x})$ are linearized at $\bm{m_X}$ to have the linear sensor model $\bm{G}=[\bm{G}_1, \bm{G}_2, \ldots, \bm{G}_M]^T$ with $\bm{G}_j=\partial g_j(\bm{m_X})/\partial \bm{x}$.
Fig. \ref{msevect} shows the MSE results for random vectors. It can be seen that the two-hop allocation using Algorithm 1 has lower MSE than all other conditions.

\section{Conclusion}
We have proposed the model for two-hop relaying wireless sensor networks and developed
an effective solution computation for joint power allocation for sensor and relay.
A consideration for nonlinear sensor networks and non-Gaussian targets is underway.

\bibliographystyle{IEEEtran}
\bibliographystyle{ieeebib}
\bibliography{GMM}

\begin{thebibliography}{10}
\providecommand{\url}[1]{#1}
\csname url@samestyle\endcsname
\providecommand{\newblock}{\relax}
\providecommand{\bibinfo}[2]{#2}
\providecommand{\BIBentrySTDinterwordspacing}{\spaceskip=0pt\relax}
\providecommand{\BIBentryALTinterwordstretchfactor}{4}
\providecommand{\BIBentryALTinterwordspacing}{\spaceskip=\fontdimen2\font plus
\BIBentryALTinterwordstretchfactor\fontdimen3\font minus
  \fontdimen4\font\relax}
\providecommand{\BIBforeignlanguage}[2]{{%
\expandafter\ifx\csname l@#1\endcsname\relax
\typeout{** WARNING: IEEEtran.bst: No hyphenation pattern has been}%
\typeout{** loaded for the language `#1'. Using the pattern for}%
\typeout{** the default language instead.}%
\else
\language=\csname l@#1\endcsname
\fi
#2}}
\providecommand{\BIBdecl}{\relax}
\BIBdecl

\bibitem{arik10}
M.~Arik and O.~Akan, ``Collaborative mobile target imaging in {UWB} wireless
  radar sensor networks,'' \emph{IEEE J. Sel. Areas Commun.}, vol.~28, no.~6,
  pp. 950--961, 2010.

\bibitem{maj09}
J.~Majchrzak, M.~Michalski, and G.~Wiczynski, ``Distance estimation with a
  long-range ultrasonic sensor system,'' \emph{IEEE Sens. J.}, vol.~9, no.~7,
  pp. 767--773, 2009.

\bibitem{nguye8}
D.~Nguyen and M.~Bagajewicz, ``Design of nonlinear sensor networks for process
  plants,'' \emph{Industrial \& Engineering Chemistry Research}, vol.~47,
  no.~15, pp. 5529--5542, 2008.

\bibitem{kar2009}
S.~Kar and J.~Moura, ``A mixed time-scale algorithm for distributed parameter
  estimation: Nonlinear observation models and imperfect communication,'' in
  \emph{Proc. IEEE Int. Conf. Acoust., Speech Signal Process. (ICASSP) 2009},
  2009, pp. 3669--3672.

\bibitem{ak02}
I.~Akyildiz, W.~Su, Y.~Sankarsubramaniam, and E.~Cayirci, ``Wireless sensor
  networks : A survey,'' \emph{Comput. Netw.}, vol.~38, pp. 393--422, 2002.

\bibitem{su08}
S.~Kim, B.~Ku, W.~Hong, and H.~Ko, ``Performance comparison of target
  localization for active sonar systems,'' \emph{IEEE Trans. Aerosp. Electron.
  Syst.}, vol.~44, no.~4, pp. 1371 --1380, 2008.

\bibitem{lian10}
L.~Liu, X.~Zhang, and H.~Ma, ``Optimal node selection for target localization
  in wireless camera sensor networks,'' \emph{IEEE Trans. Veh. Technol.},
  vol.~59, no.~7, pp. 3562 --3576, 2010.

\bibitem{ho08}
K.~Ho and L.~M. Vicente, ``Sensor allocation for source localization with
  decoupled range and bearing estimation,'' \emph{IEEE Trans. Signal Process.},
  vol.~56, no.~12, pp. 5773 --5789, 2008.

\bibitem{GV03}
M.~Gastpar and M.~Vetterli, ``Source-channel communication in sensor neworks,''
  \emph{Lecture Notes in Computer Science}, vol. 2634, pp. 162--177, 2003.

\bibitem{GV05}
------, ``Power, spatio-temporal bandwidth, and distortion in large sensor
  network,'' \emph{IEEE J. Sel. Areas Commun.}, vol.~23, no.~4, pp. 745--754,
  2005.

\bibitem{and79}
B.~Anderson and J.~Moore, \emph{Optimal Filtering}.\hskip 1em plus 0.5em minus
  0.4em\relax Englewood Cliffs, NJ: Prentice Hall, 1979.

\bibitem{shu07}
S.~Cui, J.~Xiao, A.~Goldsmith, Z.~Luo, and H.~Poor, ``Estimation diversity and
  energy efficiency in distributed sensing,'' \emph{IEEE Trans. Signal
  Process.}, vol.~55, no.~9, pp. 4683 --4695, 2007.

\bibitem{tha08}
G.~Thatte and U.~Mitra, ``Sensor selection and power allocation for distributed
  estimation in sensor networks: Beyond the star topology,'' \emph{IEEE Trans.
  Signal Process.}, vol.~56, no.~7, pp. 2649 --2661, 2008.

\bibitem{dharm8}
I.~Bahceci and A.~Khandani, ``Linear estimation of correlated data in wireless
  sensor networks with optimum power allocation and analog modulation,''
  \emph{IEEE Trans. Commun.}, vol.~56, no.~7, pp. 1146 --1156, 2008.

\bibitem{Jun09}
J.~Fang and H.~Li, ``Power constrained distributed estimation with correlated
  sensor data,'' \emph{IEEE Trans. Signal Process.}, vol.~57, no.~8, pp. 3292
  --3297, 2009.

\bibitem{rash12}
U.~Rashid, H.~D. Tuan, P.~Apkarian, and H.~H. Kha, ``Globally optimized power
  allocation in multiple sensor fusion for linear and nonlinear networks,''
  \emph{IEEE Trans. Signal Process.}, vol.~60, no.~2, pp. 903 --915, 2012.

\bibitem{R96}
T.~S. Rappaport, \emph{Wireless Communications: Principles and Practice}.\hskip
  1em plus 0.5em minus 0.4em\relax Upper Saddle River, NJ: Prentice Hall, 1996.

\end{thebibliography}

\vfill
\pagebreak
\end{document}